\documentclass{article}
\usepackage[utf8]{inputenc}
\usepackage{xcolor}
\usepackage{float}
\usepackage[justification=centering]{caption}
\usepackage[margin=0.5in]{geometry}
\usepackage[pdftex]{graphicx}
\graphicspath{{\Documents}}     
\DeclareGraphicsExtensions{.pdf,.jpeg,.png}
\usepackage{subfigure}
\usepackage{titlesec}
\usepackage{amssymb,amsmath}
\usepackage{moreverb}
\usepackage{multirow}
\usepackage[utf8]{inputenc}
\usepackage{natbib}
\usepackage{multicol}

\title{\textbf{Modern Multiple Imputation with Functional Data}}

\begin{multicols}{2}
\author{Aniruddha Rajendra Rao\\
Department of Statistics\\
\vspace{1cm}
Pennsylvania State University, USA\\
\and
Matthew Reimherr*\\
Department of Statistics\\
Pennsylvania State University, USA}
\end{multicols}

\date{}

\begin{document}

\maketitle
\vspace{-1.5cm}
\hspace{0.35cm}\textbf{Keywords:} Functional Data Analysis, Multiple Imputation, Functional Regression, Missing Data.

\abstract{This work considers the problem of fitting functional models with sparsely and irregularly sampled functional data. It overcomes the limitations of the state-of-the-art methods, which face major challenges in the fitting of more complex non-linear models. Currently, many of these models cannot be consistently estimated unless the number of observed points per curve grows sufficiently quickly with the sample size, whereas, we show numerically that a modified approach with more modern multiple imputation methods can produce better estimates in general. We also propose a new imputation approach that combines the ideas of {\it MissForest} with {\it Local Linear Forest} and compare their performance with {\it PACE} and several other multivariate multiple imputation methods. This work is motivated by a longitudinal study on smoking cessation, in which the Electronic Health Records (EHR) from Penn State PaTH to Health allow for the collection of a great deal of data, with highly variable sampling. To illustrate our approach, we explore the relation between relapse and diastolic blood pressure. We also consider a variety of simulation schemes with varying levels of sparsity to validate our methods.}

\section{Introduction}
Functional Data Analysis (FDA) is a branch of statistics that models the relationship between functions measured over a particular domain, such as time or space \citep{FDA,hor,book1,10,ramsay1997functional}
. There is a rich literature on modeling functions that are densely observed but comparatively less literature on modeling functions that are sparsely observed, which introduce new challenges. Currently, there are very few imputation methods designed for functional data \citep{10.1093/biomet/87.3.587,PMID:11252607,He2011AFM}, with a mean imputation procedure, commonly known as PACE \citep{PACE}, being the most common.

Single imputation procedures (like mean imputation or PACE) are useful in general but can't account for the uncertainty induced from the imputation procedure; once the imputation is done, analyses then typically proceeds as if the imputed values were the truth. This leads to overly optimistic measures of uncertainty and the potential for substantial bias \citep{justine}. To deal with this and other problems associated with single imputation methods, we consider multiple imputation methods.  Multiple imputation involves filling in the missing values multiple times, which creates multiple “complete” data sets.  The variability between these complete data sets reflects the uncertainty introduced in the imputation method. Multiple imputation procedures are very versatile, flexible and can be used in a wide range of settings. As multiple imputation involves creating multiple predictions for each missing value, the corresponding statistical analysis takes into account the uncertainty in the imputations and hence, yields a more reliable standard error. In simple terms, if there is less information in the observed data regarding the missing values, the imputations will be more variable, leading to higher standard errors in the analysis. In contrast, if the observed data is highly predictive of the missing values, the imputations will be more consistent across the multiple imputed data sets, resulting in  smaller and reliable standard errors \citep{10.1093/oxfordjournals.aje.a117592}.

Longitudinal studies are amenable to Functional Data Analysis (FDA), often contain sparse and irregular samples. Such data can be considered as having missing values making imputation a natural consideration. Many FDA  methods analyze fully or densely observed data sets without any appreciable missing values. However, this is often not the case when dealing with large medical and biological data. Hence, in such cases, we can either apply Sparse FDA methods \citep{PACE,FDA} or use imputation to apply more traditional FDA techniques. Several multiple imputation techniques have been proposed to impute incomplete multivariate data, including  Multivariate Imputation by Chained Equations (MICE) \citep{MICE} and MissForest (MF) \citep{Missforest}. Though these methods have not been directly applied to functional data, they have worked well in general. MICE builds a separate model for each variable (that contains missing values), conditioned on the others, which can be specified based on the data type (continuous, binary, etc.). 
MF is similar to MICE but uses random forests for building the conditional models. In both cases, variables are sequentially imputed until convergence is reached. We use the {\it MICE} \citep{micer} and {\it missForest} \citep{mfr} packages in R to implement these methods. Also, Local Linear Forest (LLF) \citep{local} which is a modification of Random Forest, is a powerful Regression method. Functional Data is naturally smooth and LLF is equipped to model signals. Taking advantage of this interesting property of LLF, we propose another imputation method similar to that of MF and MICE using LLF.

Several other imputation methods include K-nearest-neighbor (KNN) \citep{inbook}, Nonparametric imputation by data depth \citep{mozharovskyi2017nonparametric}, substantive model compatible fully conditional specification \citep{doi:10.1177/0962280214521348}, and many more. There have also been studies comparing imputation methods \citep{DING2012919,Waljeee002847,article1,Ning2012ACS} under different scenarios and data types, but for functional data, PACE has become the “gold standard”. Unfortunately, current methods for imputation in FDA are not designed to handle complex models and do not allow for consistent estimation unless one assumes that the number of observed points per curve grows sufficiently quick with the sample size. Though this is mathematically convenient, it highlights a serious concern when handling sparse functional data. Most of the methods impute while ignoring the response and subsequent modeling that is to be done with the reconstructed curves, a notable exception being Bayesian methods \citep{Kowal, Thompson}. This leads to biased estimates with unreliable standard errors and misleading p-values. For these reasons, PACE which is executed using {\it fdapace} package \citep{pacepackage}, uses an alternative approach to produce consistent estimates for functional linear models that do not generalize to non-linear models.

Missing data as described by \cite{rubin_multiple_2004}, can be divided into three categories: 1. Missing Completely at Random (MCAR), in which the missing values are independent of the observed data. 2. Missing at Random (MAR), in which the missing value patterns depend only on the observed data and are conditionally independent of the unobserved data. 3. Missing Not at Random (MNAR) also known as non-ignorable missing data or structural missing data, in which the missing data patterns depend on the observed and unobserved data. 
Usually, it is assumed that one is either working with MCAR or MAR, to make the problem tractable. We make a similar assumption in our procedure, without formally defining the missing data mechanism. Many of the recent works in Functional Data imputation \citep{articlep,articlef,articlec,He2011AFM} have built upon these ideas and adopted a missing data perspective to tackle various forms of sparsity in functional models. But all these approaches consider either a linear relationship or sparsity in the response, whereas we work with a completely observed response and sparsely observed covariates, where we can have a non-linear relation between them.

In this work, we explore the performance of several modern imputation procedures with functional data. Also, we propose another imputation method using LLF. We demonstrate how a simple modification using binning alongside careful initialization can dramatically improve 
the imputation and subsequent estimation for both linear and non-linear models. From a missing data perspective, the goal is to do imputation of the missing data in a way that retains the performance of subsequent statistical modeling.

\subsection{PaTH To Health}

Electronic Health Record (EHR) or Clinical Data often require longitudinal statistical methods, which account for the correlation between repeated measurements on the same subject. If one also assumes that these are taken from a smooth curve or data generating process, then we can exploit tools from FDA, which can produce gains in terms of flexibility and statistical power \citep{He2011AFM,article5,article6}. Since hospital visits can occur both infrequently and irregularly, we can't directly apply many FDA techniques to them. They pose a challenge to the current methods as well as for imputation. To address these challenges and illustrate the effectiveness of our approach over the current methods, we wish to apply them to an EHR data set where we predict if a smoker will relapse or not based on their Blood Pressure (BP) recordings over a span of 18 months.

The data provided by the Penn State PaTH to Health, assists research that uses patient data from multiple sources to further scientific discoveries. The data set describes patient-level data variables in a standardized manner (i.e., with the same variable name, attributes, and other metadata) along with information on demographics, encounters, diagnoses, and procedures. More information can be found on their website. 

While there is a wealth of research related to smoking and blood pressure (BP) \citep{articles1,articles2,articles3}, our goal here is not to make deep scientific statements, but rather to illustrate the utility of our methods, which we hope will prove useful to practitioners.  The highest risk of relapse for smokers is during the end of their first and second year after quitting \citep{article7,article8}. We, therefore, focus on modeling the relapse of the patients based on monitoring their BP within the first two years, which may be useful to the practitioners designing interventions for patients at risk of relapse.

\begin{figure}[h]
\begin{center}
\centerline{\includegraphics[height=7cm]{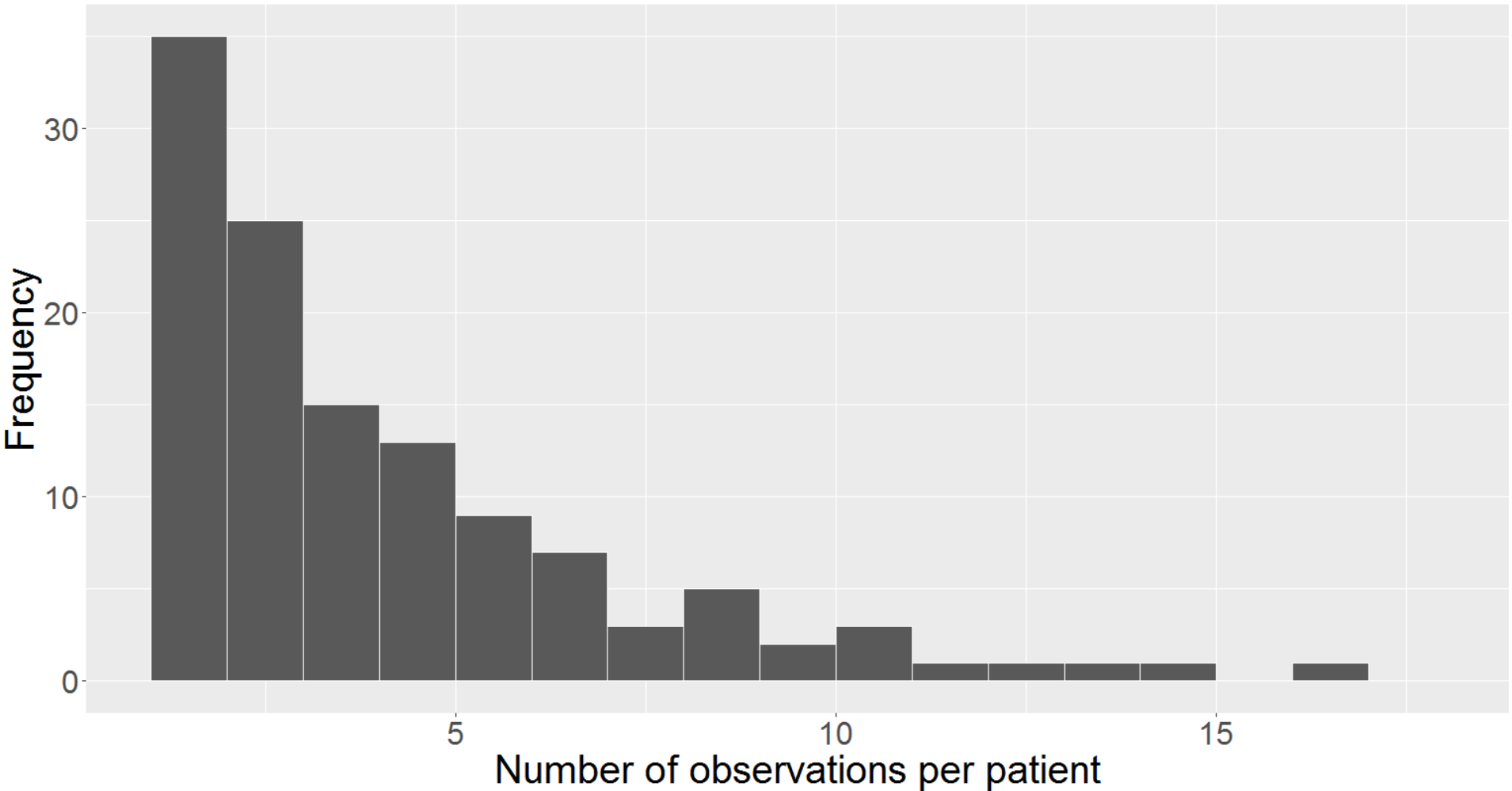}}
\caption{Histogram of the number of observations for BP per patient, ranging from 1 to 18.}
\label{fig: patient}
\end{center}
\end{figure}

\begin{figure}[h]
\begin{center}
\centerline{\includegraphics[height=7cm]{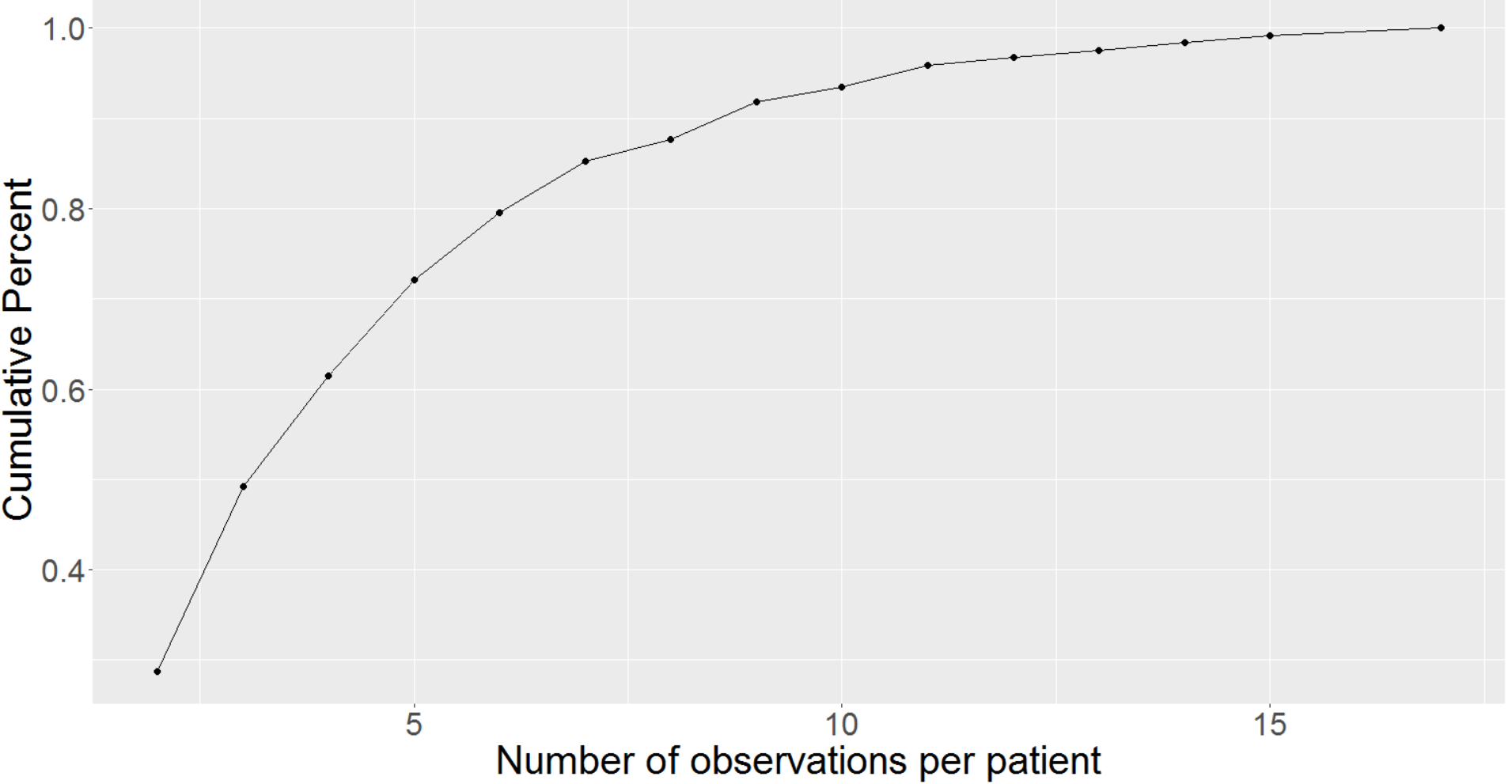}}
\caption{Cumulative percentage of observations for BP per patient.}
\label{fig: patient1}
\end{center}
\end{figure}

In general, EHR data sets vary significantly with the timing and regularity of the appointments, and clinical measurements are affected by errors of varying types and degrees \citep{10.1093/jamia/ocx037}. Similar challenges apply to measurements in the PaTH data set where we have various kinds of clinical measurements and information recorded. The ability to characterize trajectories of sparse irregular data has potential applicability to many clinical questions. Though the term sparsity is somewhat subjective in the context of functional/longitudinal data, many of the patients in the Path data set have just 2 measurements, while the most number of measurements for any patient is 17 (after cleaning and implementing the exclusion criteria). We can see from Figure \ref{fig: patient} the modal number of measurements is 2, while relatively few patients had more than 7 clinical visits and almost none had more than 11. Also, from the cumulative observation Figure (appendix Figure \ref{fig: patient1}), we observe that 94\% of the patients had 10 or fewer measurements, 72\% had at most 5 measurements, and 28\% had no more than 2 measurements. On an average, we have around 4 measurements per patient. Sparsity arises due to many reasons in an EHR setting. Some patients never come back, some patients are not observed with any uniformity or regularity, etc. Having identified the BP trajectories as both sparse and irregular, we move to introduce the imputation methods that account for these conditions in a Functional framework before revisiting this data in Section 3.

\subsection{Organization}
The rest of the paper is organized as follows. In Section 2, we briefly go through PACE and Multivariate Imputation methods (MICE and MF) before introducing our proposed method using LLF, and modifications to the Multivariate Imputation methods using bins and careful initialization, to better deal with Functional Data. We present multiple simulations for the linear and non-linear cases under different values of sparsity in Section 3. This section also includes the EHR data, where we fit a scalar-on-function regression model to determine if a patient (smoker) will relapse or not at the end of 18 months using BP as the functional predictor. These examples help us to illustrate the limitations of previous approaches and demonstrate the usefulness of our methodology, which overcomes many of the issues discussed earlier. In the final section, we present our concluding remarks and future research directions, which pertain to better understanding of the bins and deeper statistical theory.

\section{Methods}

In this section, we give details of the current imputation methods for the scalar-on-function regression model. In subsection 2.1 we define the necessary notation used in the paper. In subsection 2.2, we briefly discuss scalar-on-function regression models. Subsection 2.3 gives an overview of PACE and the multivariate imputation method MICE in detail along with their shortcomings. In subsection 2.4, we discuss LLF and the multivariate imputation method MF. We present our new imputation procedure that extends the ideas of MF to LLF, as well as discuss how to use careful binning and initialization to improve performance.

\subsection{Setup and Notation}
We assume the data is collected from trajectories that are independent realizations of a smooth random function, with unknown mean function $E(X(t))$=$\mu(t)$ and covariance function $C_X(t,s):=cov(X(s),X(t))$. We define the underlying functional covariates as $\left\{X_{i}(t): t \in[0,1] ; 1 \leq i \leq n\right\},$ where $t$ denotes the argument of the functions, usually time, and $i$ denotes the subject or unit.  We assume that these curves are only observed at times $t_{i j}$ ($j=1, \ldots, m_{i}$) with some error:
$$
x_{i j}=X_{i}\left(t_{i j}\right)+\delta_{i j}
$$
Let $\mathbf{x}_{i}=\left(x_{i 1}, \ldots, x_{i m}\right)^{\top}$ denote the vector of observed values on the function $X_{i} .$ Let $Y_{i}$, be the outcome, which is a function of $X_{i}$ and some error. Examples of such relations can be found in the next subsection.

Generally, when integration is written without limits, it is implied to be over the entire domain, usually standardized to $[0,1]$ for simplicity. The main focus of this work is to develop tools for consistently estimating the parameters in functional regression models.

\subsection{Functional models}

The scalar-on-function regression model in the linear case is defined as:
\begin{equation} \label{1}
Y_{i}=\alpha+\int \beta(t) X_{i}(t) dt+\varepsilon_{i}
\end{equation}
A common way of estimating the model components is by Basis or Functional Principal Component (FPC) expansions, where we simplify the problem of estimating the parameters by projecting the functions to a finite dimension and then using multiple regression and least squares \citep{FDA,ramsay1997functional}.

Non-linear modeling becomes very challenging with functional data due to {\it the curse of dimensionality}. One popular way of simplifying the problem is by using the  generalized additive model, also known as continuously additive model \citep{McLean,articlel1,articlel2,articlel3,articlel4}:
\begin{equation} \label{3}
Y_i=\int f(X_i(t),t)dt + \epsilon_i
\end{equation}
where the bivariate function $(x,t) 	\rightarrow f(x,t)$ is smooth, but unknown. It is commonly estimated using either basis expansions \citep{PACE, Greven} (often with a tensor product basis) or using Reproducing Kernel Hilbert Spaces \citep{articlel2, reimherr2017optimal}. \cite{McLean} developed Functional Generalized Additive Models that enabled non-scalar response mapping.

Most methods for fitting the models discussed above require densely sampled functional data.  For irregular and sparsely sampled data that are observed with error, estimating the modeling parameters becomes much more challenging. Directly smoothing the $x_{i j}$ to plug into a dense estimation framework seems like a straightforward idea but can result in substantial bias.

\subsection{Imputation Methods}

PACE \citep{PACE} uses functional principal components
(FPC) analysis, in which the FPC scores are imputed using conditional expectations. In addition to the requirements discussed previously, PACE also relies heavily on the data being Gaussian. The   Karhunen-Loève or Principal Component expansion of $X_{i}(t)$ is given by:
\begin{equation} \label{2}
X_{i}(t)=\mu_{X}(t)+\sum_{j=1}^{\infty} \xi_{i j} v_{j}(t),
\end{equation}
where $v_{j}(t)$ are the eigenfunctions of $C_{X}$ with eigenvalues $\lambda_{1} \geq \lambda_{2} \geq \cdots \geq 0$. The scores are computed as:
\begin{equation} \label{4}
    \xi_{i j}=\left\langle X_{i}-\mu_{X}, v_{j}\right\rangle
\end{equation}
PACE proceeds by computing the conditional expectation of the scores given the observed data. This conditioning method is straightforward and tends to work much better than direct smoothing of $x_{i j}$. It provides the Best Linear Unbiased Predictors (BLUPs) under Gaussian assumptions and works in the presence of both measurement errors and sparsity. We can plug in the BLUPs values into a dense estimation framework to model the response.

PACE still suffers from a few major problems. One issue is that the imputation procedure of PACE does not consider the response $Y_{i}$ nor does it have any consideration for subsequent models that will be fit. This results in a bias while estimating model parameters \citep{justine}. In addition, PACE is just a single imputation method and hence the uncertainty in the imputation is not properly propagated when forming confidence intervals, prediction intervals, or p-values.  For this reason, the PACE software \citep{pacepackage} uses an alternative approach for fitting linear models, which does not extend to non-linear models.

After understanding PACE, we now look into some of the standard methods used for imputation in the multivariate case. All of these methods have proven to work well in the multivariate setting but have never been tested in the functional setting, where the sample sparsity can be very high.

Multivariate Imputation by Chained Equations (MICE) \citep{MICE}, also known as “fully conditional specification” or “sequential regression multiple imputation”, has emerged in the statistical literature as one of the principal methods for addressing missing data. MICE performs multiple imputations rather than single imputation and hence, it can account for the statistical uncertainty. In addition, the chained equations approach is very flexible and can handle variables of varying types (e.g. continuous or categorical) as well as complexities such as bounds or survey skip patterns.  At a high level, the MICE procedure is a series of models whereby each variable with missing data is modeled conditional upon the other variables in the data. The MICE procedure is as follows: 1. We start by initialization, wherein we fill in all the missing values with mean imputation. 2. Next, we select the first variable with missing entries. A model is fit with this variable as the outcome and the other variables as predictors. 3. The missing values of the current variable are then replaced with predicted (imputed) values from the model in step 2. 4. Step 2 and step 3 are repeated while rotating through the variables with missing values sequentially. The cycling through each of the variables constitutes one iteration or cycle. 5. At the end of one cycle, all of the missing values have been replaced with predictions from the model that reflect the relationships observed in the data. The cycles are repeated a few times and after each cycle the imputed values are updated.

The number of cycles to be performed is pre-specified and after the last cycle, the final imputations are retained, resulting in one imputed data set. Generally, 10-15 cycles are performed. The idea is that, by the end of the cycles, the distribution of the parameters governing the imputations (e.g. the coefficients in the regression models) should have converged, in the sense of becoming stable. Different MICE software packages vary somewhat in their exact implementation of this algorithm (e.g. in the order in which the variables are imputed), but the general strategy is the same. Here, we have used the  {\it MICE} \citep{micer} package in R.

A key advantage of MICE is its flexibility in using different models. Generally, the modeling techniques included in MICE are Predictive Mean Matching, Linear Regression, Generalized Linear Models, Bayesian Methods, Random Forest, Linear Discriminant Analysis, and many more. Its primary disadvantage is that it does not have the same  theoretical justification as other imputation methods. In particular, fitting a series of conditional distributions, as is done using the series of regression models, may not be consistent with proper joint distribution, though some research suggests that this may not be a large issue in applied settings \citep{article3}.

\subsection{Our Approach}
\subsubsection{MissForest and Local Linear Forest}

MissForest (MF) \citep{Missforest} is a multiple imputation method, which proceeds by training a Random Forest (RF) on the observed parts of the data. Random forest \citep{10.1023/A:1010933404324} is a non-parametric method that is able to deal with mixed data types as well as allow for interactive and non-linear effects. MF addresses the missing data problem using an iterative imputation scheme by training a RF on observed values in the first step, followed by predicting the missing values in the next step and then proceeding iteratively. RF works well in High dimensional cases with good accuracy and robustness. Though the idea of MF is similar to MICE, they differ in the ordering scheme of the columns to be imputed, and MICE requires certain assumptions about the distribution of the data or subsets of the variables, which may or may not be true.

For an arbitrary variable $p$ in $X_{n \times m}$ $(p=1,2,...m)$ including missing values at entries $\mathbf{i}^{(p)} \operatorname{mis} \subseteq\{1, \dots, n\}$, we can separate the data set into four parts: The observed values for variable $p$, denoted by $\mathbf{y}^{(p)}_{obs}$, the missing values for variable $p$, denoted by $\mathbf{y}^{(p)}_{mis}$, the variables other than $p$ with observations at $\mathbf{i}^{(p)} \operatorname{obs} \subseteq\{1, \dots, n\}$ / $\mathbf{i}^{(p)}_{mis}$ denoted by $\mathbf{x}^{(p)}_{obs}$ and the variables other than $p$ with observations at $\mathbf{i}^{(p)}_{mis}$ denoted by $\mathbf{x}^{(p)}_{mis}$.

The approach is as follows: We initialize for the missing values in $X$ using mean imputation or another imputation method. We then sort the variables $p$ in $X$ ($p=1,...,m$) in ascending order of the missing values. For each variable $p$, the missing values are imputed by first fitting an Random Forest with response $\mathbf{y}^{(p)}_{obs}$ and predictors $\mathbf{x}^{(p)}_{obs}$; then, predicting the missing values $\mathbf{y}^{(p)}_{mis}$ by applying the trained RF to $\mathbf{x}^{(p)}_{mis}$. We sequentially do this for all variables with missing values, that is one cycle. The imputation procedure is repeated for multiple cycles until a stopping criterion is met.

The advantage of MF is that it can deal with any kind of data. Also, MF is straightforward, as it does not need any tuning of parameters nor does it require any assumption about distributional aspects of the data. The full potential of MF is deployed when the data includes complex interactions or non-linear relations between variables of different types, which is not possible with PACE. Furthermore, MF can be applied to high-dimensional data sets with a low sample size and still provide excellent results. MF often outperforms other methods in terms of imputation \citep{Missforest}, but the method has no smoothing mechanism and hence the imputed values of the curves are not smooth. To deal with this and to increase model accuracy, we integrate binning into the method as explained in the next subsection.

Local Linear Forest \citep{local} uses a RF to generate weights that are used as a kernel for local linear regression, i.e., Local Linear Forest takes the RF weights $\alpha_{i}\left(\mathbf{x}_{i}\right)$ and uses them to solve:

\begin{equation}\label{eq:111}
    \min _{\mu, \beta} \sum_{i=1}^{n}\left(Y_{i}-\mu-\left(x-x_{i}\right)^{\prime} \beta\right)^{2}\alpha_{i}\left(\mathbf{x}_{i}\right)
\end{equation}

The RF weights $\alpha_{i}\left(\mathbf{x}_{i}\right)$ are found with the help of the leaf $L_b(x_i)$ in each tree $T_b$ in a forest of $B$ trees as follows:
\begin{equation}\label{eq:11}
  \alpha_{i}\left(\mathbf{x}_{i}\right)=\frac{1}{B} \sum_{b=1}^{B} \frac{1\left\{\mathbf{X}_{\mathbf{i}} \in L_{b}\left(\mathbf{x}_{i}\right)\right\}}{\left|L_{b}\left(\mathbf{x}_{i}\right)\right|}  
\end{equation}
where $\sum_{i=1}^{n} \alpha_{i}\left(\mathbf{x}_{i}\right)=1$ and for each $i, 0 \leq \alpha_{i}\left(\mathbf{x}_{i}\right) \leq 1 .$ \cite{athey2016generalized} used this perspective to harness RF for solving weighted estimating equations, and gave asymptotic guarantees on the resulting predictions. With the help of the above weights, LLF solves the locally weighted least squares problem.

LLF is a modification of Local Linear Regression with the help of RF, equipped to model signals and fix bias issues. We use this to our advantage and propose a modification to the MF method, where we replace the RF with LLF. We refer to this as Miss Local Linear Forest (MLLF). This new approach using LLF for imputation follows the same steps as MF but instead of using RF as the modeling technique to impute the missing values, we will now use LLF. Since LLF does not inherit the multiple imputation like MF, we generate multiple imputed sets and take an average like in the case of MICE. MLLF has similar benefits to MF. We update the MissForest code \citep{mfr} in R using {\it grf} package \citep{grf} to implement MLLF.

\subsubsection{Adapting methods to Functional data}

One of the key features of Functional Data is the smoothness of the underlying curves. MF and MLLF produce well-imputed curves but are not smooth. We overcome this problem with the help of binning and careful initialization. We improve the initialization by using PACE instead of simple mean imputation. These boosted methods using PACE are denoted as MFP for MissForest and MLLFP for Local Linear Forest. While this leads to higher performance in general with slightly smoother imputed curves, it does not directly smooth the imputed curves or resulting model parameters. Also, this initialization comes with a computational burden as PACE itself is computationally heavy. Another restriction which all of these imputation methods have, except PACE, is that they need to pass through the observed points, which need not be optimal, especially in the presence of observation noise.

We overcome the non-smoothness issue and computational problem by the use of bins. Binning (also known as Discrete binning or bucketing) is a data pre-processing method that is used to reduce the effects of minor observation errors and smooth the data. The original data values which fall in a given small interval, a bin, are replaced by a value representative for that interval, usually the mean value. Binning aggregates the values into a fixed range. So, we divide the desired grid of the time points into $k$ bins and impute over the $k$ points before interpolating back to $m$ time points using b-splines. Here, $k$ is a tuning parameter that acts much like a bandwidth in kernel smoothing. This not only leads to smoother imputation results but also improves the subsequent modeling as well. As binning helps in reducing the number of time points ($m<k$), the overall process becomes way more computational friendly. The way bins are defined is as follows:
\begin{itemize}
    \item The first bin is the first time point of the data.
    \item The last bin is the last time point of the data.
    \item The middle $(k$-$2)$ bins are divided into equal parts and are represented by the mean of all values in that bin.
\end{itemize}

We denote the methods with the binning as MF\_B for MissForest and MLLF\_B for Local Linear Forest. If we add PACE initialization to it then we denote the methods as MFP\_B for MissForest and MLLFP\_B for Local Linear Forest. We can see in the next section that, this not only leads to much smoother results but also improves imputations and modeling performance.

\section{Simulation and Results}
Throughout this section, we refer to MissForest as MF, MissForest with PACE initialization as MFP, binning without PACE as MF\_B and with PACE as MFP\_B. Miss Local Linear Forest (MLLF) uses an analogous naming scheme. In addition to comparing all the methods in simulations, we will compare the results for the EHR data as well. In the simulation, we compare them in both Linear and Non-Linear scalar-on-function regression settings with a scalar and binary response, investigating the imputation accuracy, model fit (prediction accuracy), and $\beta$ estimates (only for the Linear case).  We compare across multiple simulated data sets with varying time points observed $m$, sample sizes $n$ and sparsity $s$.

\subsection{Simulation}
\textbf{Linear case:}

For the Linear case, we simulate $n$ iid random curves $\left\{X_{1}(t), \cdots, X_{n}(t)\right\}$ from a Gaussian process with mean 0 and covariance
$$
  C_{X}(t, s)=\frac{\sigma^{2}}{\Gamma(\nu) 2^{\nu-1}}\left(\frac{\sqrt{2 \nu}|t-s|}{\rho}\right)^{\nu} K_{\nu}\left(\frac{\sqrt{2 \nu}|t-s|}{\rho}\right),
$$
which is the Matérn covariance function and $K_{\nu}$ is the modified Bessel function of the second kind. We set $\rho=0.5,$ $\nu=5 / 2$ and $\sigma^{2}=1 .$ These curves are evaluated at $m$ equally-spaced time points from $[0,1]$. We assume that each observed point contains a normal measurement error with mean zero and variance $\sigma_{\delta}^{2}=0.3 .$ We set $\beta(t)=w \times \sin (2 \pi t),$ where $w$ is a weight coefficient used to adjust the signal. The response, $Y_{i}$ $(i=1, \cdots, n)$ is computed using the model in Equation \eqref{1}, where $\alpha=0$ and $\sigma_{\epsilon}^{2}=1 .$ In the binary response case, we define $Y_{i}$ $(i=1, \cdots, n)$ using Bernoulli and logit link function in Equation \eqref{1}. Finally, for each curve, we assume a percentage ($s$) of the $m$ time points are unobserved. After the scores are imputed, we fit a scalar-on-function regression model using these imputed curves.

For the linear case, we simulate the data sets of different sample sizes, $n \in\{200,500,1000\}$ (results for n=200 and n=1000 are included in the appendix); different number of observations per curve, $m \in\{32,52\}$; and with different sparsity level, $s \in\{Medium, High\}$. For sparsity levels, Medium means $50\%$ of the points are missing for each curve and High means more than $85\%$ of the points are missing for each curve. Also, the values of $m$ are taken such that it helps with the process of binning. Each of these settings is simulated 10 times. Since we are primarily interested in the accuracy of the final estimates $\hat{\beta}(t)$,  $\hat{Y}(t)$ and $\hat{X}(t),$ we report the Root Mean Square Error (RMSE) or Prediction Error for each of them.

\begin{table}[h]
\centering
\footnotesize
\begin{tabular}{|c|c|c|c|c|c|c|c|c|c|c|c|c|}
\hline
\multicolumn{1}{|c|}{Method} & \multicolumn{6}{c|}{n=500, s=Medium}                                                                  & \multicolumn{6}{c|}{n=500, s=High}                                                                  \\ \cline{2-13} 
                        & \multicolumn{3}{c|}{m=32, b=17}                  & \multicolumn{3}{c|}{m=52, b=27}                  & \multicolumn{3}{c|}{m=32, b=8}                   & \multicolumn{3}{c|}{m=52, b=12}                  \\ \cline{2-13} 
                        & Pred           & $\beta$              & Imp            & Pred           & $\beta$              & Imp            & Pred           & $\beta$              & Imp            & Pred           & $\beta$              & Imp            \\ \hline
MF                      & 0.136          & 0.155          & 0.108          & 0.227          & 0.241          & \textbf{0.077} & 0.267          & 0.422          & 0.470           & 0.177          & 0.307          & 0.390           \\ \hline
PACE                    & 0.170           & 0.208          & 0.199          & 0.484          & 0.595          & 1.910           & 0.376          & 0.432          & \textbf{0.369} & 0.350          & 0.388          & 0.308          \\ \hline
MLLF                    & 0.142          & \textbf{0.149} & \textbf{0.070}  & 0.234          & 0.242          & \textbf{0.023} & 58.010          & 43.721           & 0.601          & 69.24          & 55.43          & 0.508          \\ \hline
MICE                    & 3.611          & 3.612          & 0.090           & 0.237          & 0.253          & 0.089          & 9.840           & 5.950           & 0.910           & 3.162           & 2.390           & 1.073          \\ \hline
MFP                     & \textbf{0.122} & \textbf{0.144} & 0.105          & 0.228          & 0.250          & 0.130           & 0.141          & \textbf{0.289} & \textbf{0.398} & 0.232          & 0.310           & 0.328          \\ \hline
MLLFP                   & 0.132          & 0.153          & 0.144          & 0.232          & 0.246          & 0.100           & 0.376          & 0.432          & \textbf{0.364} & 0.384          & 0.386          & 0.302          \\ \hline
MF\_B                   & \textbf{0.122} & \textbf{0.136} & \textbf{0.079} & \textbf{0.173} & \textbf{0.180} & \textbf{0.052} & \textbf{0.117} & \textbf{0.264} & \textbf{0.334} & \textbf{0.152} & \textbf{0.217} & \textbf{0.261} \\ \hline
MLLF\_B                 & \textbf{0.126} & \textbf{0.137} & \textbf{0.082} & \textbf{0.174} & \textbf{0.177} & \textbf{0.053} & \textbf{0.097} & \textbf{0.291} & \textbf{0.338} & \textbf{0.132} & \textbf{0.203} & \textbf{0.267} \\ \hline
MFP\_B                  & \textbf{0.122} & \textbf{0.138} & \textbf{0.053} & \textbf{0.176} & \textbf{0.182} & \textbf{0.023} & \textbf{0.101} & \textbf{0.263} & \textbf{0.339} & \textbf{0.146} & \textbf{0.219} & \textbf{0.275} \\ \hline
MLLFP\_B                & \textbf{0.128} & \textbf{0.143} & \textbf{0.059} & \textbf{0.179} & \textbf{0.178} & \textbf{0.023} & \textbf{0.091} & 0.817          & 0.614          & \textbf{0.129} & \textbf{0.214} & 0.307          \\ \hline
\end{tabular}
\caption{\label{tab:a2} RMSE of Prediction, $\beta$ coefficients and Imputation of the curves for different methods\\ under Linear case when n=500 for different time points and sparsity settings.}
\end{table}

\begin{table}[h]
\centering
\footnotesize
\begin{tabular}{|c|c|c|c|c|c|c|c|c|c|c|c|c|}
\hline
\multicolumn{1}{|c|}{Method} & \multicolumn{6}{c|}{n=500, s=Medium}                                                                  & \multicolumn{6}{c|}{n=500, s=High}                                                                  \\ \cline{2-13} 
                        & \multicolumn{3}{c|}{m=32, b=17}                  & \multicolumn{3}{c|}{m=52, b=7}                  & \multicolumn{3}{c|}{m=32, b=17}                   & \multicolumn{3}{c|}{m=52, b=12}                  \\ \cline{2-13} 
                        & Pred           & $\beta$              & Imp            & Pred           & $\beta$              & Imp            & Pred           & $\beta$              & Imp            & Pred           & $\beta$              & Imp            \\ \hline
MF                      & 0.268          & 0.409          & \textbf{0.043}          & \textbf{0.280}          & 0.391          & {0.155} & 0.317         & 0.383          & 0.187           & \textbf{0.251}          & 0.347          & 0.307           \\ \hline
PACE                    & 0.270           & 0.508          & 0.199          & 0.431          & 0.426          & 1.910           & 0.348          & 0.463          & {0.369} & 0.466           &  0.687          & 0.308          \\ \hline
MLLF                    & 0.364          & {0.453} & \textbf{0.040}  & 0.532          &  1.376          & {0.212} & 0.282          & 0.490          & 0.309          & 0.374          & 2.768          & 1.715         \\ \hline
MICE                    & 0.459          & 1.282          & 0.284           & 0.589          & 0.385          & 0.934          & 0.543           & 4.828           & 1.115           & 0.672           & 1.814           & 1.050          \\ \hline
MFP                     & {0.260} & {0.385} & 0.198          & \textbf{0.282}          & 0.340          & 0.198           & 0.258          & {0.356} & {0.185} & \textbf{0.260}          & 0.360           & \textbf{0.282}      \\ \hline
MLLFP                   & 0.274          & 0.417          & \textbf{0.038}          & 0.364          & 0.379          & 0.213           & 0.360          & 0.462          & {0.307} & 0.366          & 0.372          & 0.651      \\ \hline
MF\_B                   & \textbf{0.160} & \textbf{0.262} & \textbf{0.036} &  \textbf{0.276} & \textbf{0.322} & \textbf{0.132} & \textbf{0.214} & {0.329} & \textbf{0.174} & \textbf{0.250} & \textbf{0.290} & \textbf{0.261} \\ \hline
MLLF\_B                 & \textbf{0.161} & \textbf{0.235} & \textbf{0.037}  & \textbf{0.264} & {0.378} & \textbf{0.132} & \textbf{0.232} & {0.346} & \textbf{0.169} &\textbf{0.246} & \textbf{0.271} & \textbf{0.274} \\ \hline
MFP\_B                  & \textbf{0.161} & \textbf{0.249} & \textbf{0.037} & \textbf{0.266} & \textbf{0.321} & \textbf{0.121} &  \textbf{0.220} & \textbf{0.309} & \textbf{0.179} & \textbf{0.244} & \textbf{0.296} & \textbf{0.275} \\ \hline
MLLFP\_B                & \textbf{0.169} & \textbf{0.270} & \textbf{0.036} &  \textbf{0.274} & {0.357}          & \textbf{0.121} & \textbf{0.236} & {0.467} & {0.250}         & \textbf{0.252} & \textbf{0.281} & 0.304          \\ \hline
\end{tabular}
\caption{\label{tab:ab2} Prediction Error, RMSE of $\beta$ coefficients and RMSE of Imputation of the curves for different methods\\ under Linear case with Binary response when n=500 for different time points and sparsity settings.}
\end{table}

\begin{figure}[h]
\begin{center}
\centerline{\includegraphics[height=8cm]{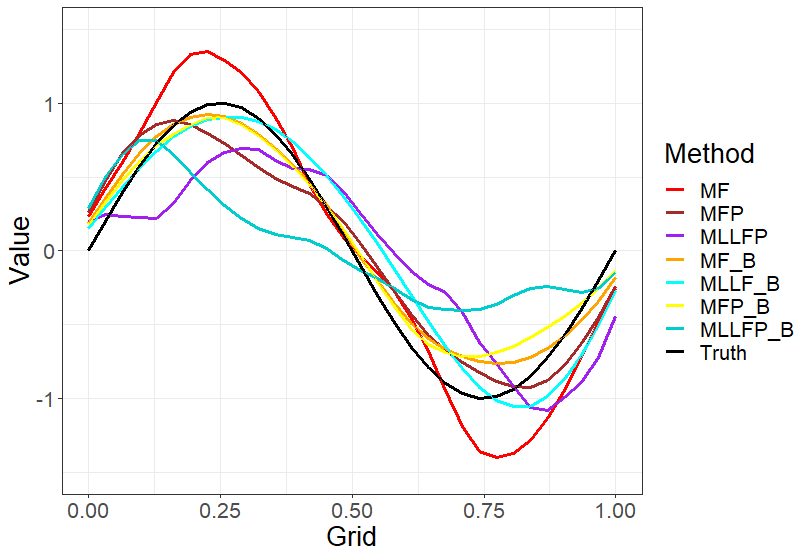}}
\centering
\caption{Estimated coefficient function for different methods under linear case with sample size\\ (n)=500, time points (m)=52, sparsity (s)=High and scalar response.}
\label{fig:a}
\end{center}
\end{figure}

Table \ref{tab:a2} and Table \ref{tab:ab2} indicates that, in general, irrespective of the number of points, all the binned methods perform better compared to other methods for Prediction Error or RMSE of Prediction, $\beta$ coefficients and Imputation when the sparsity is Medium and the sample size is 500. Also, we can see that the RMSE of imputation for PACE is the same for both the tables. This is because we are using the same sample curves to generate scalar and binary response.
There is no clear winner between MF and MLLF within the binned methods with or without PACE initialization. Again, for High sparsity, we notice a similar behavior as before, irrespective of the number of points, all the binned methods perform better. As we increase the sparsity, it seems like MICE and MLLF perform worse. This happens mainly because they do a poor job of imputing the curves, the effects of which get compounded when estimating the parameters and modeling. Again, binned methods are the best with no clear winner. The results for the other cases with different sample sizes and scalar response yield similar performance to these tables and can be found in the appendix Table \ref{tab:a1} for $n=200$ and appendix Table \ref{tab:a3} for $n=1000$.

We can see from Figure \ref{fig:a}, how each method does in estimating the $\beta$ coefficient. We can observe that most of the MissForest extensions are catching the right shape and doing a good job. The plot does not contain PACE, MICE and MLLF as their estimates were very poor, which is also reflected in the RMSE for the $\beta$ coefficients from Table \ref{tab:a2}.

\begin{figure}[h]
\begin{center}
\centerline{\includegraphics[height=8cm]{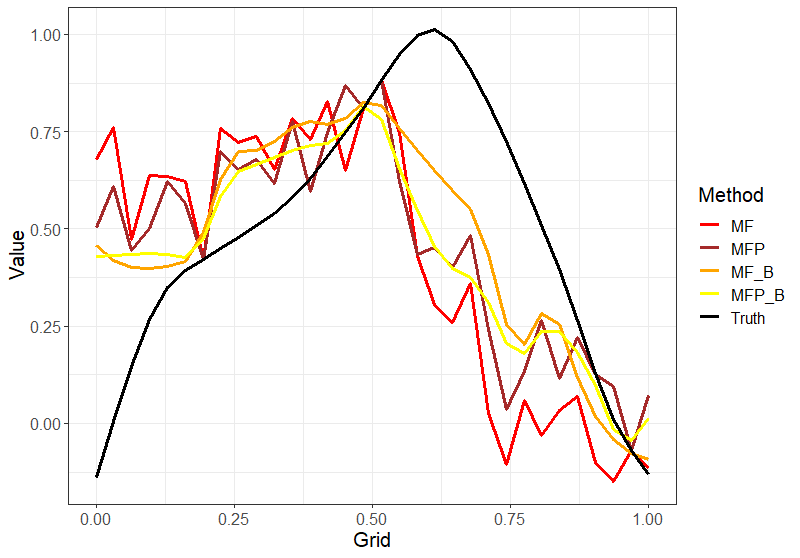}}
\caption{Comparing imputed curves in non-binned and binned methods of MF under linear case\\ for one random sample curve with time points (m)=52 and sparsity (s)=High.}
\label{fig:b}
\end{center}
\end{figure}

Figure \ref{fig:b} shows an example of imputed curves under different methods for one random sample curve. Here, the main focus is to see how binning helps in not only doing better imputation as seen from Table \ref{tab:a2} but also the fact that it gives much smoother results as compared to the MissForest methods without the binning. The same effect can be seen with Local Linear Forest methods as well. This plot is included in the appendix Figure \ref{fig:q1}.

\noindent\textbf{Non-Linear case:}

All simulations parameters are the same as before, except the response $Y_{i}$ $(i=1, \cdots, n)$ is computed using the model in Equation \eqref{3}, where $f(X_i(t),t)=5*sin(X(t)^2*t^2)$. 
For the Non-linear case, we also simulate data sets of sample size $n$ as 500 with different number of observations per curve, $m \in\{32,52\}$; and with different sparsity, $s \in\{Medium, High\} .$ Each of these settings is simulated 10 times. Since we are primarily interested in the accuracy of the final output $\hat{Y}(t)$ and $\hat{X}(t),$ we report the Root Mean Square Error (RMSE) or Prediction Error for each of them. Another Non-Linear model result can be found in the appendix Table\ref{tab:a5}.

\begin{table}[h]
\centering
\begin{tabular}{|c|c|c|c|c|c|c|c|c|}
\hline
\multicolumn{1}{|c|}{Method} & \multicolumn{4}{c|}{n=500, s=Medium}                                & \multicolumn{4}{c|}{n=500, s=High}                                \\ \cline{2-9} 
                        & \multicolumn{2}{c|}{m=32, b=17} & \multicolumn{2}{c|}{m=52, b=27} & \multicolumn{2}{c|}{m=32, b=8}  & \multicolumn{2}{c|}{m=52, b=12} \\ \cline{2-9} 
                        & Pred           & Imp            & Pred           & Imp            & Pred           & Imp            & Pred           & Imp            \\ \hline
MF                      & 0.236          & 0.102          & 0.223          & 0.078          & 0.303          & 0.428          & 0.355          & 0.381          \\ \hline
PACE                    & 0.328          & 0.238          & 0.214          & 1.253          & 0.431          & 0.659          & 0.386          & 0.551          \\ \hline
MLLF                    & 0.230           & 0.066          & 0.209          & 0.064          & 0.419          & 0.592          & 0.351          & 0.482          \\ \hline
MICE                    & 0.334          & 0.812          & 0.214          & 0.680           & 0.652          & 0.892          & 0.644          & 1.020           \\ \hline
MFP                     & 0.235          & 0.980           & 0.210           & 0.075          & 0.334          & 0.379          & 0.351          & 0.324          \\ \hline
MLLFP                   & 0.276          & \textbf{0.044} & 0.560           & \textbf{0.027} & 0.427          & 0.357          & 0.342          & \textbf{0.259} \\ \hline
MF\_B                   & \textbf{0.174} & \textbf{0.045} & \textbf{0.161} & \textbf{0.053} & \textbf{0.293} & \textbf{0.257} & \textbf{0.318} & \textbf{0.275} \\ \hline
MLLF\_B                 & \textbf{0.183} & \textbf{0.045} & \textbf{0.163} & \textbf{0.054} & \textbf{0.335} & \textbf{0.364} & \textbf{0.312} & \textbf{0.282} \\ \hline
MFP\_B                  & \textbf{0.176} & \textbf{0.031} & \textbf{0.160}  & \textbf{0.026} & \textbf{0.290}  & \textbf{0.257} & \textbf{0.311} & \textbf{0.251} \\ \hline
MLLFP\_B                & \textbf{0.174} & \textbf{0.031} & \textbf{0.163} & \textbf{0.028} & \textbf{0.328} & 0.385          & \textbf{0.329} & 0.308          \\ \hline
\end{tabular}

\caption{\label{tab:a4} RMSE of Prediction and Imputation of the curves for different methods under Non-Linear case $(f(X_i(t),t)=5*sin(X(t)^2*t^2))$ when n=500 for different time points and sparsity settings.}

\end{table}

\begin{table}[h]
\centering
\begin{tabular}{|c|c|c|c|c|c|c|c|c|}
\hline
\multicolumn{1}{|c|}{Method} & \multicolumn{4}{c|}{n=500, s=Medium}                                & \multicolumn{4}{c|}{n=500, s=High}                                \\ \cline{2-9} 
                        & \multicolumn{2}{c|}{m=32, b=17} & \multicolumn{2}{c|}{m=52, b=12} & \multicolumn{2}{c|}{m=32, b=17}  & \multicolumn{2}{c|}{m=52, b=12} \\ \cline{2-9} 
                        & Pred           & Imp            & Pred           & Imp            & Pred           & Imp            & Pred           & Imp            \\ \hline
MF                      & 0.388          & 0.157          & 0.390          & 0.296          & 0.478          & 0.522          & 0.306          & 0.402          \\ \hline
PACE                    & 0.405          & 0.238          & 0.356          & 1.253          & 0.589          & 0.659          & 0.411          & 0.551          \\ \hline
MLLF                    & 0.386          & 0.144          & 0.392          & 0.274          & 0.486          & 2.790          & 0.382          & 1.490          \\ \hline
MICE                    & 0.451          & 0.154          & 0.388          & {0.282}           & 0.641          & 1.262          & 0.712          & 2.139           \\ \hline
MFP                     & 0.388          & 0.213          & {0.292}           & 0.238          & 0.492          & 0.573          & \textbf{0.300}          & 0.708          \\ \hline
MLLFP                   & 0.382          & {0.268} & {0.288}           & {0.276} &  \textbf{0.290}    & 0.413          & 0.402          & 1.401        \\ \hline
MF\_B                   & \textbf{0.296} & \textbf{0.112} & \textbf{0.262} & \textbf{0.232} & \textbf{0.308} & \textbf{0.374} & \textbf{0.288} & \textbf{0.372} \\ \hline
MLLF\_B                 & \textbf{0.294} & \textbf{0.113} & \textbf{0.262} & \textbf{0.232} & \textbf{0.294} & \textbf{0.369} & \textbf{0.292} & \textbf{0.372} \\ \hline
MFP\_B                  & \textbf{0.294} & \textbf{0.091} & \textbf{0.262}  & \textbf{0.221} & \textbf{0.302}  & \textbf{0.379} & \textbf{0.290} & \textbf{0.366} \\ \hline
MLLFP\_B                & \textbf{0.294} & \textbf{0.090} & \textbf{0.262} & \textbf{0.221} & \textbf{0.280} & 0.550          & \textbf{0.292} & \textbf{0.370}          \\ \hline
\end{tabular}

\caption{\label{tab:ab4} Prediction Error and RMSE of the curves for different methods under Non-Linear case with\\ Binary response $(f(X_i(t),t)=5*sin(X(t)^2*t^2))$ when n=500 for different time points and sparsity settings.}

\end{table}

From Table \ref{tab:a4} and Table \ref{tab:ab4}, we observe that our proposed approaches are outperforming PACE and MICE for Imputation irrespective of the number of points and sparsity. When it comes to prediction, our methods are still better than PACE and MICE but the gap is not as large compared to the linear case. Overall, we see the same trend as in the Linear case, with the binned methods outperforming every other method under various simulation settings.

The major takeaway from all the simulations is that our methods perform the best under various settings. This is because our methods impute smoother curves resulting in better modeling and smoother estimates of the beta coefficients irrespective of the relation between the response and the functional predictors.

\subsection{EHR}

New statistical tools are vital for data such as PATH, which are very large and have a great deal of underlying structure. We see the performance of the developed tools for imputation with this longitudinal/functional data. The electronic medical records contain information about smokers (patients) who irregularly come for a check-up at the hospital. We have the blood pressure (BP) readings along with some other measurement values at each check-up of the patients. From previous studies, we know that majority of the relapse among smokers occur within the first two years. For cleaning the data, the exclusion criteria were based on the number of longitudinal measurements (time points).  Patients who had a smoking history (at least smoked for a year) with less than 2 measurements were excluded. After cleaning, we are left with 122 patients, of which 61 smokers relapsed and 61 smokers didn't, where each smoker is under observation for 18 months. Hence, here the sample size (n) is 122 and the number of time points (m) is 18. The data is sparse naturally as the patients don't come in for check-up regularly and sometimes the measurements are missing even though there was a visit due to unknown reasons. We build a model to predict if a patient will relapse or not using the Blood Pressure (BP) measurement over an 18 month period.

\begin{table}[h]
\centering
\begin{tabular}{|l|l|l|l|l|l|l|l|l|l|l|}
\hline
Method     & MF & PACE & MLLF & MICE & MFP & MLLFP & MF\_B & MLLF\_B & MFP\_B  & MLLFP\_B \\ \hline
Linear Model   &  0.39 & 0.42 & 0.39 & 0.39 & 0.36 & 0.37 & \textbf{0.33} & 0.35 & \textbf{0.32} & 0.35\\ \hline
CAM   &  0.36 & 0.39 & 0.36 & 0.38 & 0.35 & 0.36 & \textbf{0.28} & 0.32 & \textbf{0.28} & 0.31\\ \hline

\end{tabular}
\caption{\label{tab:e} Prediction error of different methods for the EHR data.}
\end{table}

\begin{figure}[h]
\begin{center}
\centerline{\includegraphics[height=7cm]{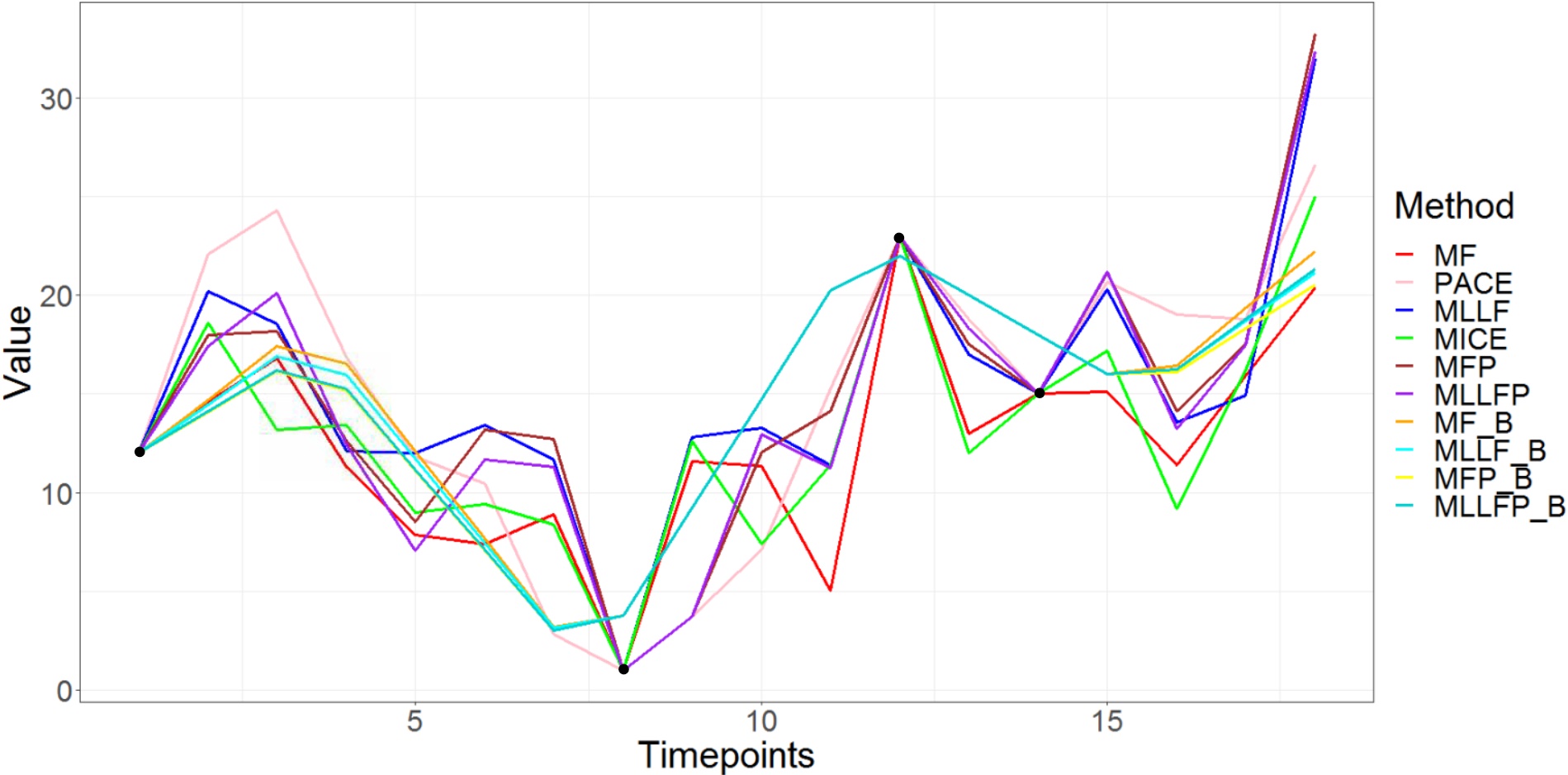}}
\caption{Imputation results for one curve from the EHR data.}
\label{fig:f}
\end{center}
\end{figure}

\begin{figure}[h]
\begin{center}
\centerline{\includegraphics[height=7cm]{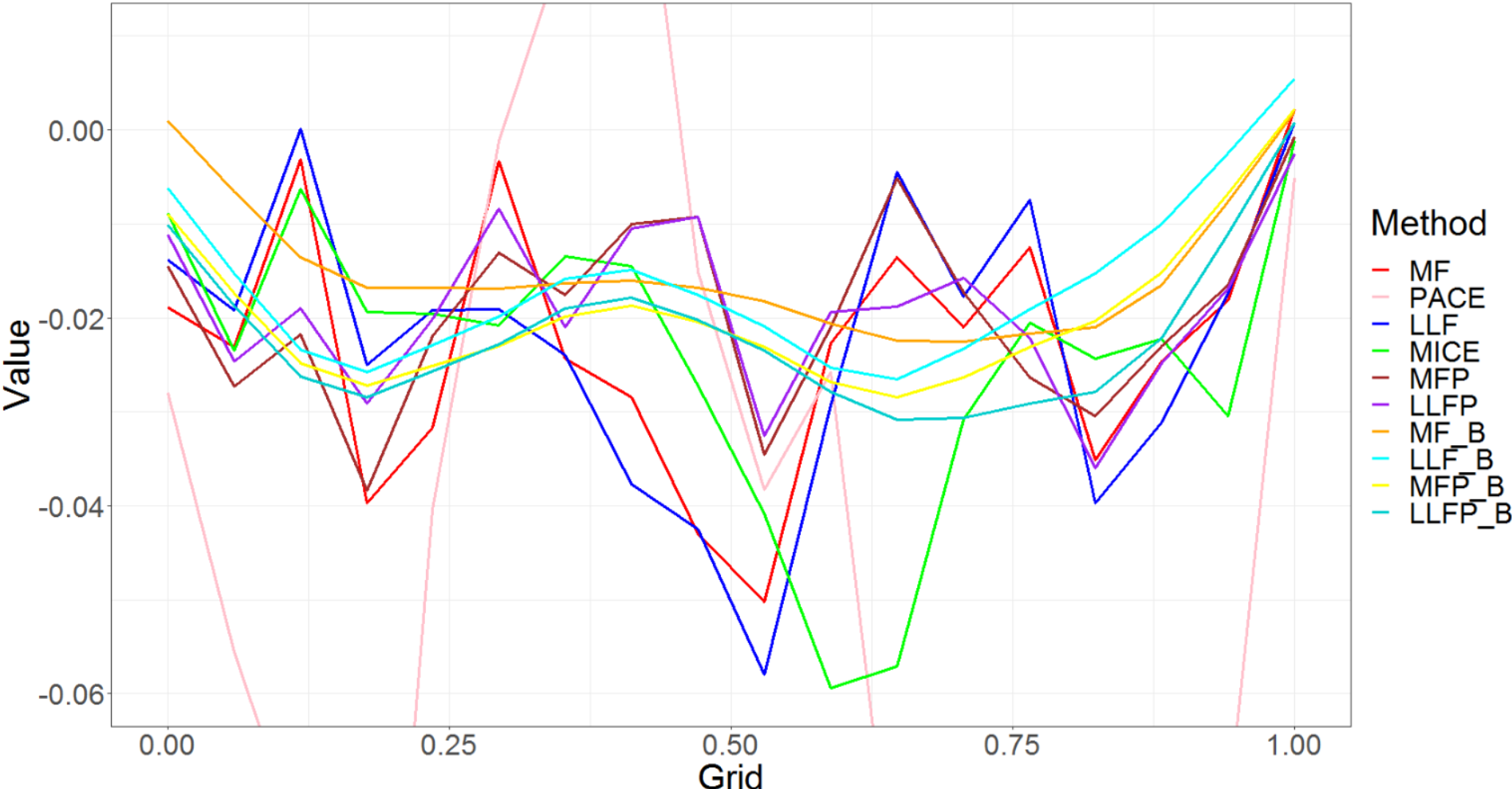}}
\caption{Estimated coefficient function of all the methods for the EHR data using\\ linear scalar-on-function regression.}
\label{fig:e}
\end{center}
\end{figure}

We can infer from Table \ref{tab:e} that the binned methods again are outperforming the other methods in prediction irrespective of linear or non-linear modeling. Also, since both the model results are so close, the true relation looks linear. This is further supported by Figure \ref{fig:f}, where we see only the binned methods have smooth imputed curves with the black points denoting the observed values for that patient. Although all the estimated $\beta$ coefficients seem to follow the same trend in Figure \ref{fig:e}, the methods with binning have much smoother results, leading to better interpretability. Smoothness is inherent to functional data and that is why binned methods are able to perform so well.

Also, Figure \ref{fig:e} suggests that patients with low BP or sudden changes in their BP have a higher risk of relapse. Also, the curve isn't constant suggesting that acceleration/velocity of the BP curve is important. We did check and found out that the average BP was higher in the control group (no relapse) than for the cases (relapse). We feel there might be confounding variables and further analysis is needed, which is out of the scope of the project as we are only interested in demonstrating the efficacy of our methods for imputation and training the model, which results in better analysis and interpretation.

\section{Discussion and Conclusions}
In this project, we explored different multivariate imputation methods under sparse and irregular functional data setting. We have proposed a new imputation method Missing Local Linear Forest (MLLF) which is a mixture of MissForest (MF) and MICE. Also, we modified this method along with MF to deal with Functional Data in a systematic fashion by careful initialization using PACE and smoothing out the results using Bins. Our proposed approaches overcome a lot of the challenges faced by the current methods (like PACE and MICE) to give consistent estimates. They incorporate the response as well as deal with complex non-linear relations with multiple imputations. Results under multiple simulation settings also illustrated the value of our approach over existing methods for fitting scalar-on-function regression models when the functional predictors are irregularly and sparsely sampled irrespective of the sparsity level, number of points in the curve and sample size. All the binned methods work equally well with slight variations in some cases; though there is no clear winner, MissForest with binning (MF\_B) was the most consistent performer. 

Our approach is sensitive to the subdivision of the time points into the bins. Different binning strategies were not explored in depth but are one of the directions for further investigation. Another interesting avenue is defining a relationship between time points ($m$) and the number of bins ($k$), to ease the search for the optimum bin number. Also, even though it looks like the extension of MissForest (MF) performs better than Local Linear Forest imputation (MLLF), further analysis is required to differentiate between the methods. Deep learning has become a major research area in multiple fields and the application of Neural Networks into the imputation setting might be very interesting though our initial efforts did not bear strong results.

Finally, at a high level, there are still many remaining challenges with the imputation of functional data.  When evaluating the performance of future methods, we suggest considering at least three critical points: (1) does the imputations improve subsequent modeling, (2) can the imputations incorporate the assumed underlying smoothness of the curves or at least domain information, and (3) can the imputations handle measurement noise in the observed points?  A multiple imputation approach seems to be critical for the first point, while the latter two are still quite open.  Our binning approach, while simple, helped a great deal with the second point.  However, the third point was basically untouched in this work.  When using methods such as PACE, incorporating observation error is straightforward, but it is unclear how to incorporate it into more complicated imputation procedures.

\section*{Acknowledgement}
This research was supported in part by the following grant to Pennsylvania State University:  NSF SES-1853209

\bibliographystyle{apalike}
\bibliography{ref}

\begin{thebibliography}{}

\bibitem[Acuna and Rodriguez, 2004]{inbook}
Acuna, E. and Rodriguez, C. (2004).
\newblock The treatment of missing values and its effect on classifier
  accuracy.
\newblock {\em Journal of Classification}, pages 639--647.

\bibitem[Athey et~al., 2016]{athey2016generalized}
Athey, S., Tibshirani, J., and Wager, S. (2016).
\newblock Generalized random forests.

\bibitem[Bartlett et~al., 2015]{doi:10.1177/0962280214521348}
Bartlett, J.~W., Seaman, S.~R., White, I.~R., Carpenter, J.~R., and for~the
  Alzheimer's Disease Neuroimaging~Initiative* (2015).
\newblock Multiple imputation of covariates by fully conditional specification:
  Accommodating the substantive model.
\newblock {\em Statistical Methods in Medical Research}, 24(4):462--487.
\newblock PMID: 24525487.

\bibitem[Breiman, 2001]{10.1023/A:1010933404324}
Breiman, L. (2001).
\newblock Random forests.
\newblock {\em Mach. Learn.}, 45(1):5–32.

\bibitem[Buuren and Groothuis-Oudshoorn, 2011]{micer}
Buuren, S. and Groothuis-Oudshoorn, C. (2011).
\newblock Mice: Multivariate imputation by chained equations in r.
\newblock {\em Journal of Statistical Software}, 45.

\bibitem[Carnahan-Craig et~al., 2018]{article5}
Carnahan-Craig, S., Blankenberg, D., Parodi, A., Paul, I., Birch, L., Savage,
  J., Marini, M., Stokes, J., Nekrutenko, A., Reimherr, M., Chiaromonte, F.,
  and Makova, K. (2018).
\newblock Child weight gain trajectories linked to oral microbiota composition.
\newblock {\em Scientific Reports}, 8.

\bibitem[Chen et~al., 2019]{pacepackage}
Chen, Y., Carroll, C., Dai, X., Fan, J., Hadjipantelis, P.~Z., Han, K., Ji, H.,
  Mueller, H.-G., and Wang, J.-L. (2019).
\newblock {\em fdapace: Functional Data Analysis and Empirical Dynamics}.
\newblock R package version 0.5.1.

\bibitem[Crambes and Henchiri, 2018]{articlec}
Crambes, C. and Henchiri, Y. (2018).
\newblock Regression imputation in the functional linear model with missing
  values in the response.
\newblock {\em Journal of Statistical Planning and Inference}, 201.

\bibitem[Daymont et~al., 2017]{10.1093/jamia/ocx037}
Daymont, C., Ross, M.~E., Russell~Localio, A., Fiks, A.~G., Wasserman, R.~C.,
  and Grundmeier, R.~W. (2017).
\newblock {Automated identification of implausible values in growth data from
  pediatric electronic health records}.
\newblock {\em Journal of the American Medical Informatics Association},
  24(6):1080--1087.

\bibitem[Ding and Ross, 2012]{DING2012919}
Ding, Y. and Ross, A. (2012).
\newblock A comparison of imputation methods for handling missing scores in
  biometric fusion.
\newblock {\em Pattern Recognition}, 45(3):919 -- 933.

\bibitem[Fan et~al., 2015]{articlel3}
Fan, Y., James, G.~M., and Radchenko, P. (2015).
\newblock Functional additive regression.
\newblock {\em Ann. Statist.}, 43(5):2296--2325.

\bibitem[Ferraty and Romain, 2011]{book1}
Ferraty, F. and Romain, Y. (2011).
\newblock {\em The Oxford Handbook of Functional Data Analysis}.
\newblock Oxford University Press.

\bibitem[Ferraty et~al., 2012]{articlef}
Ferraty, F., Sued, M., and Vieu, P. (2012).
\newblock Mean estimation with data missing at random for functional
  covariables.
\newblock {\em Statistics}, iFirst.

\bibitem[Ferraty and Vieu, 2006]{10}
Ferraty, F. and Vieu, P. (2006).
\newblock {\em Nonparametric Functional Data Analysis: Theory and Practice
  (Springer Series in Statistics)}.
\newblock Springer-Verlag, Berlin, Heidelberg.

\bibitem[Friedberg et~al., 2018]{local}
Friedberg, R., Tibshirani, J., Athey, S., and Wager, S. (2018).
\newblock Local linear forests.

\bibitem[García-Rodríguez et~al., 2013]{article8}
García-Rodríguez, O., Secades-Villa, R., Florez-Salamanca, L., Okuda, M.,
  Liu, S.-M., and Blanco, C. (2013).
\newblock Probability and predictors of relapse to smoking: Results of the
  national epidemiologic survey on alcohol and related conditions (nesarc).
\newblock {\em Drug and alcohol dependence}, 132.

\bibitem[Goldsmith and Schwartz, 2017]{article6}
Goldsmith, J. and Schwartz, J. (2017).
\newblock Variable selection in the functional linear concurrent model.
\newblock {\em Statistics in medicine}, 36.

\bibitem[Greenland and Finkle, 1995]{10.1093/oxfordjournals.aje.a117592}
Greenland, S. and Finkle, W.~D. (1995).
\newblock {A Critical Look at Methods for Handling Missing Covariates in
  Epidemiologic Regression Analyses}.
\newblock {\em American Journal of Epidemiology}, 142(12):1255--1264.

\bibitem[Greven et~al., 2010]{Greven}
Greven, S., Crainiceanu, C., Caffo, B., and Reich, D. (2010).
\newblock Longitudinal functional principal component analysis.
\newblock {\em Electronic journal of statistics}, 4:1022--1054.

\bibitem[Hansson et~al., 1996]{articles2}
Hansson, L., Hedner, T., and Jern, S. (1996).
\newblock Smoking affects blood pressure.
\newblock {\em Blood pressure}, 5:68.

\bibitem[He et~al., 2011]{He2011AFM}
He, Y., Yucel, R.~M., and Raghunathan, T.~E. (2011).
\newblock A functional multiple imputation approach to incomplete longitudinal
  data.
\newblock {\em Statistics in medicine}, 30 10:1137--56.

\bibitem[Herd et~al., 2009]{article7}
Herd, N., Borland, R., and Hyland, A. (2009).
\newblock Predictors of smoking relapse by duration of abstinence: findings
  from the international tobacco control (itc) four country survey.
\newblock {\em Addiction (Abingdon, England)}, 104:2088--99.

\bibitem[Horv{\'a}th and Kokoszka, 2012]{hor}
Horv{\'a}th, L. and Kokoszka, P. (2012).
\newblock {\em Inference for Functional Data with Applications}.
\newblock Springer Series in Statistics. Springer New York.

\bibitem[James et~al., 2000]{10.1093/biomet/87.3.587}
James, G., Hastie, T., and Sugar, C. (2000).
\newblock {Principal component models for sparse functional data}.
\newblock {\em Biometrika}, 87(3):587--602.

\bibitem[Kokoszka and Reimherr, 2018]{FDA}
Kokoszka, P. and Reimherr, M. (2018).
\newblock {\em Introduction to Functional Data Analysis}.
\newblock New York: Chapman and Hall/CRC.

\bibitem[Kowal et~al., 2019]{Kowal}
Kowal, D.~R., Matteson, D.~S., and Ruppert, D. (2019).
\newblock Functional autoregression for sparsely sampled data.
\newblock {\em Journal of Business \& Economic Statistics}, 37(1):97--109.

\bibitem[Liao et~al., 2014]{article1}
Liao, S., Lin, Y., Kang, D., Chandra, D., Bon, J., Kaminski, N., Sciurba, F.,
  and Tseng, G. (2014).
\newblock Missing value imputation in high-dimensional phenomic data: Imputable
  or not, and how?
\newblock {\em BMC bioinformatics}, 15:346.

\bibitem[Ma and Zhu, 2016]{articlel4}
Ma, H. and Zhu, Z. (2016).
\newblock Continuously dynamic additive models for functional data.
\newblock {\em Journal of Multivariate Analysis}, 150:1 -- 13.

\bibitem[McLean et~al., 2014]{McLean}
McLean, M.~W., Hooker, G., Staicu, A.-M., Scheipl, F., and Ruppert, D. (2014).
\newblock Functional generalized additive models.
\newblock {\em Journal of Computational and Graphical Statistics},
  23(1):249--269.

\bibitem[Mozharovskyi et~al., 2017]{mozharovskyi2017nonparametric}
Mozharovskyi, P., Josse, J., and Husson, F. (2017).
\newblock Nonparametric imputation by data depth.

\bibitem[MÃ¼ller et~al., 2013]{articlel1}
MÃ¼ller, H.-G., Wu, Y., and Yao, F. (2013).
\newblock {Continuously additive models for nonlinear functional regression}.
\newblock {\em Biometrika}, 100(3):607--622.

\bibitem[Ning and Cheng, 2012]{Ning2012ACS}
Ning, J. and Cheng, P.~E. (2012).
\newblock A comparison study of nonparametric imputation methods.
\newblock {\em Statistics and Computing}, 22:273--285.

\bibitem[Petrovich et~al., 2018]{justine}
Petrovich, J., Reimherr, M., and Daymont, C. (2018).
\newblock Highly irregular functional generalized linear regression with
  electronic health records.

\bibitem[Preda et~al., 2010]{articlep}
Preda, C., Saporta, G., and Mbarek, M. (2010).
\newblock The nipals algorithm for missing functional data.
\newblock {\em Revue Roumaine de Mathématiques Pures et Appliquées}, 55.

\bibitem[Primatesta et~al., 2001]{articles3}
Primatesta, P., Falaschetti, E., Gupta, S., Marmot, M., and Poulter, N. (2001).
\newblock Association between smoking and blood pressure : Evidence from the
  health survey for england.
\newblock {\em Hypertension}, 37:187--93.

\bibitem[Ramsay and Silverman, 1997]{ramsay1997functional}
Ramsay, J. and Silverman, B. (1997).
\newblock {\em Functional Data Analysis}.
\newblock Springer series in statistics. Springer.

\bibitem[Reimherr et~al., 2017]{reimherr2017optimal}
Reimherr, M., Sriperumbudur, B., and Taoufik, B. (2017).
\newblock Optimal prediction for additive function-on-function regression.

\bibitem[Rice and Wu, 2001]{PMID:11252607}
Rice, J. and Wu, C. (2001).
\newblock Nonparametric mixed effects models for unequally sampled noisy
  curves.
\newblock {\em Biometrics}, 57(1):253—259.

\bibitem[Rubin, 2004]{rubin_multiple_2004}
Rubin, D. (2004).
\newblock {\em Multiple imputation for nonresponse in surveys}.
\newblock Wiley classics library edition. Wiley, Hoboken, {NJ}. [u.a.].

\bibitem[Schafer and Graham, 2002]{article3}
Schafer, J. and Graham, J. (2002).
\newblock Missing data: Our view of the state of the art.
\newblock {\em Psychological Methods}, 7:147--177.

\bibitem[Stekhoven, 2013]{mfr}
Stekhoven, D.~J. (2013).
\newblock {\em missForest: Nonparametric Missing Value Imputation using Random
  Forest}.
\newblock R package version 1.4.

\bibitem[Stekhoven and Bühlmann, 2011]{Missforest}
Stekhoven, D.~J. and Bühlmann, P. (2011).
\newblock Missforest-non-parametric missing value imputation for mixed-type
  data.
\newblock {\em Bioinformatics}, 28(1):112--118.

\bibitem[Thompson and Rosen, 2008]{Thompson}
Thompson, W. and Rosen, O. (2008).
\newblock A bayesian model for sparse functional data.
\newblock {\em Biometrics}, 64:54--63.

\bibitem[Tibshirani et~al., 2020]{grf}
Tibshirani, J., Athey, S., and Wager, S. (2020).
\newblock {\em grf: Generalized Random Forests}.
\newblock R package version 1.1.0.

\bibitem[van Buuren, 2007]{MICE}
van Buuren, S. (2007).
\newblock Multiple imputation of discrete and continuous data by fully
  conditional specification.
\newblock {\em Statistical Methods in Medical Research}, 16(3):219--242.
\newblock PMID: 17621469.

\bibitem[Waljee et~al., 2013]{Waljeee002847}
Waljee, A.~K., Mukherjee, A., Singal, A.~G., Zhang, Y., Warren, J., Balis, U.,
  Marrero, J., Zhu, J., and Higgins, P.~D. (2013).
\newblock Comparison of imputation methods for missing laboratory data in
  medicine.
\newblock {\em BMJ Open}, 3(8).

\bibitem[Wang and Ruppert, 2013]{articlel2}
Wang, X. and Ruppert, D. (2013).
\newblock Optimal prediction in an additive functional model.
\newblock {\em Statistica Sinica}, 25.

\bibitem[Wang et~al., 2018]{articles1}
Wang, Y., Zheng, X., Zhang, C., Yang, Y., Liu, L., Qi, Y., and Bu, P. (2018).
\newblock A12426 association between smoking and blood pressure in elderly male
  patients with essential hypertension.
\newblock {\em Journal of Hypertension}, 36:e321.

\bibitem[Yao et~al., 2005]{PACE}
Yao, F., Müller, H.-G., and Wang, J.-L. (2005).
\newblock Functional data analysis for sparse longitudinal data.
\newblock {\em Journal of the American Statistical Association},
  100(470):577--590.

\end{thebibliography}

\newpage
\section{Appendix}

\begin{figure}[h]
\begin{center}
\centerline{\includegraphics[height=7cm]{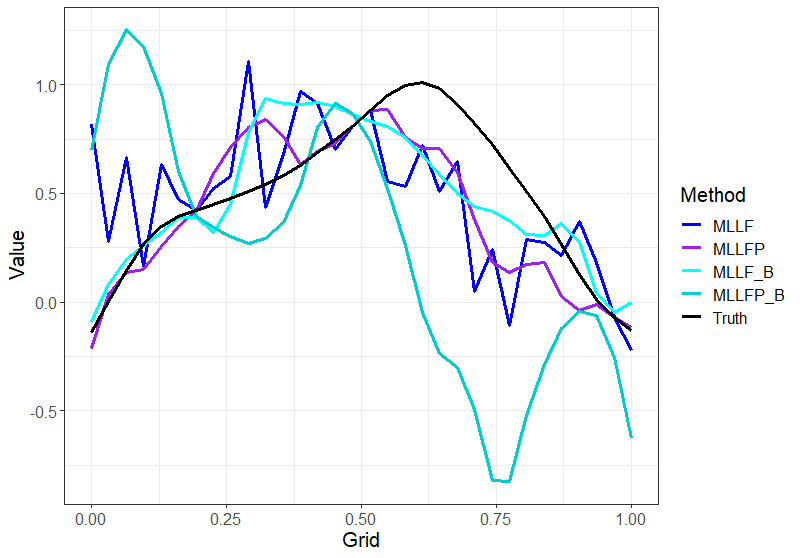}}
\caption{Comparing imputed curves in non-binned and binned methods of MLLF under linear case\\ for one random sample curve with time points (m)=52 and sparsity (s)=High.}
\label{fig:q1}
\end{center}
\end{figure}

\begin{table}[h] 
\centering
\footnotesize
\begin{tabular}{|c|c|c|c|c|c|c|c|c|c|c|c|c|}
\hline
\multicolumn{1}{|c|}{Method}& \multicolumn{6}{c|}{n=200, s=Medium}                                                                  & \multicolumn{6}{c|}{n=200, s=High}                                                                  \\ \cline{2-13} 
                        & \multicolumn{3}{c|}{m=32, b=17}                  & \multicolumn{3}{c|}{m=52, b=27}                  & \multicolumn{3}{c|}{m=32, b=8}                   & \multicolumn{3}{c|}{m=52, b=12}                  \\ \cline{2-13} 
                        & Pred           & $\beta$              & Imp            & Pred           & $\beta$             & Imp            & Pred           & $\beta$              & Imp            & Pred           & $\beta$              & Imp            \\ \hline
MF                      & 0.149          & 0.1656         & 0.141          & 0.476          & 0.459          & 0.117          & 1.290          & 1.272          & 0.573          & 0.349          & 0.476          & 0.529          \\ \hline
PACE                    & 0.221          & 0.332          & 0.393          & 0.700          & 0.667          & 1.863          & 0.433          & 0.461          & 0.434          & 0.449          & 0.456          & 0.362          \\ \hline
MLLF                    & 0.140          & 0.144          & \textbf{0.085} & 0.515          & 0.491          & \textbf{0.025} & 30.871         & 20.224         & 0.663          & 26.841         & 20.301         & 0.583          \\ \hline
MICE                    & 0.148          & 0.191          & 0.204          & 0.462          & 0.452          & 0.172          & 0.400          & 0.704          & 0.939          & 0.440          & 0.649          & 0.961          \\ \hline
MFP                     & 0.140          & 0.158          & 0.165          & 0.468          & 0.455          & 0.225          & 0.340          & 0.483          & 0.509          & 0.220          & 0.354          & 0.451          \\ \hline
MLLFP                   & 0.170          & 0.188          & 0.186          & 0.493          & 0.472          & 0.196          & 0.434          & 0.461          & 0.427          & 0.454          & 0.459          & 0.354          \\ \hline
MF\_B                   & \textbf{0.124} & \textbf{0.135} & \textbf{0.89}  & \textbf{0.264} & \textbf{0.258} & 0.078          & \textbf{0.285} & \textbf{0.340} & \textbf{0.395} & \textbf{0.191} & \textbf{0.255} & \textbf{0.312} \\ \hline
MLLF\_B                 & \textbf{0.124} & \textbf{0.132} & \textbf{0.091} & \textbf{0.266} & \textbf{0.260} & 0.077          & \textbf{0.286} & \textbf{0.313} & 0.434          & \textbf{0.178} & \textbf{0.283} & \textbf{0.301} \\ \hline
MFP\_B                  & \textbf{0.126} & \textbf{0.133} & \textbf{0.071} & \textbf{0.266} & \textbf{0.269} & \textbf{0.037} & 0.411          & 0.434          & 0.450          & 0.211          & \textbf{0.262} & \textbf{0.293} \\ \hline
MLLFP\_B                & \textbf{0.124} & \textbf{0.135} & \textbf{0.088} & \textbf{0.268} & \textbf{0.265} & \textbf{0.030} & 0.423          & 0.662          & 0.652          & \textbf{0.197} & \textbf{0.273} & 0.345          \\ \hline
\end{tabular}
\caption{\label{tab:a1} RMSE of Prediction, $\beta$ coefficients and Imputation of the curves for different methods under\\ Linear case when $n=200$ under different time points and sparsity settings.}
\end{table}


\begin{table}[h]
\centering
\footnotesize
\begin{tabular}{|c|c|c|c|c|c|c|c|c|c|c|c|c|}
\hline
\multicolumn{1}{|c|}{Method} & \multicolumn{6}{c|}{n=1000, s=Medium}                                                                 & \multicolumn{6}{c|}{n=1000, s=High}                                                                 \\ \cline{2-13} 
                        & \multicolumn{3}{c|}{m=32, b=17}                  & \multicolumn{3}{c|}{m=52, b=27}                  & \multicolumn{3}{c|}{m=32, b=7}                   & \multicolumn{3}{c|}{m=52, b=27}                  \\ \cline{2-13} 
                        & Pred           & $\beta$              & Imp            & Pred           & $\beta$              & Imp            & Pred           & $\beta$              & Imp            & Pred           & $\beta$              & Imp            \\ \hline
MF                      & 0.155          & \textbf{0.163} & \textbf{0.085} & 0.259          & 0.265          & \textbf{0.061} & 3.137          & 3.070          & 0.404          & 0.395          & 0.455          & 0.328          \\ \hline
PACE                    & 0.176          & 0.242          & 0.192          & \textbf{0.125} & 0.583          & 0.868          & 0.412          & 0.441          & \textbf{0.389} & 0.326          & 0.354          & 0.281          \\ \hline
MLLF                    & \textbf{0.140} & \textbf{0.144} & \textbf{0.061} & 0.265          & 0.272          & \textbf{0.027} & 51.947         & 41.243         & 0.545          & 61.500         & 51.082         & 0.476          \\ \hline
MICE                    & 0.152          & 0.168          & 0.118          & 0.254          & 0.262          & 0.092          & 0.365          & 0.660          & 0.873          & 0.911          & 0.838          & 0.856          \\ \hline
MFP                     & 0.151          & \textbf{0.164} & \textbf{0.085} & 0.251          & 0.278          & 0.110          & 0.198          & \textbf{0.296} & \textbf{0.353} & 0.288          & 0.338          & \textbf{0.274} \\ \hline
MLLFP                   & 0.167          & 0.198          & 0.134          & 0.0.255        & 0.284          & 0.099          & 0.412          & 0.441          & \textbf{0.339} & 0.324          & 0.357          & \textbf{0.244} \\ \hline
MF\_B                   & \textbf{0.143} & \textbf{0.168} & \textbf{0.071} & \textbf{0.116} & \textbf{0.220} & \textbf{0.046} & \textbf{0.149} & \textbf{0.267} & \textbf{0.357} & \textbf{0.104} & \textbf{0.194} & \textbf{0.263} \\ \hline
MLLF\_B                 & \textbf{0.147} & \textbf{0.153} & \textbf{0.072} & \textbf{0.113} & \textbf{0.208} & \textbf{0.046} & \textbf{0.123} & \textbf{0.293} & \textbf{0.373} & \textbf{0.111} & \textbf{0.265} & \textbf{0.295} \\ \hline
MFP\_B                  & \textbf{0.144} & \textbf{0.165} & \textbf{0.048} & \textbf{0.116} & \textbf{0.221} & \textbf{0.026} & \textbf{0.145} & \textbf{0.273} & \textbf{0.348} & \textbf{0.128} & \textbf{0.219} & 0.366          \\ \hline
MLLFP\_B                & \textbf{0.143} & \textbf{0.160} & \textbf{0.051} & \textbf{0.113} & \textbf{0.211} & \textbf{0.031} & \textbf{0.127} & 0.802          & 0.644          & \textbf{0.138} & 0.354          & 0.475          \\ \hline
\end{tabular}
\caption{\label{tab:a3} RMSE of Prediction, $\beta$ coefficients and Imputation of the curves for different methods under\\ Linear case when $n=1000$ under different time points and sparsity settings.}
\end{table}

\begin{table}[h] 
\centering
\begin{tabular}{|c|c|c|c|c|c|c|c|c|} 
\hline
\multicolumn{1}{|c|}{Method} & \multicolumn{4}{c|}{n=500, s=Medium}                                & \multicolumn{4}{c|}{n=500, s=High}                                \\ \cline{2-9} 
                        & \multicolumn{2}{c|}{m=32, b=17} & \multicolumn{2}{c|}{m=52, b=12} & \multicolumn{2}{c|}{m=32, b=7}  & \multicolumn{2}{c|}{m=52, b=27} \\ \cline{2-9} 
                        & Pred           & Imp            & Pred           & Imp            & Pred           & Imp            & Pred           & Imp            \\ \hline
MF                      & 0.180          & 0.189          & 0.292          & 0.112          & 0.682          & 0.541          & 0.832          & 0.468          \\ \hline
PACE                    & 0.147          & 1.912          & 0.253          & 0.549          & 0.598          & 0.393          & 0.557          & 0.292          \\ \hline
MLLF                    & 0.138          & 0.104          & 0.239          & 0.093          & 25.376         & 2.586          & 18.231         & 1.581          \\ \hline
MICE                    & 0.199          & 0.275          & 0.448          & 0.267          & 0.712          & 0.837          & 0.802          & 1.007          \\ \hline
MFP                     & 0.178          & 0.177          & 0.237          & 0.102          & 0.630          & 0.434          & 0.842          & 0.431          \\ \hline
MLLFP                   & 0.320          & 0.060          & 0.212          & 0.091          & 0.483          & 0.320          & 0.570          & \textbf{0.160} \\ \hline
MF\_B                   & \textbf{0.155} & \textbf{0.054} & \textbf{0.189} & \textbf{0.032} & \textbf{0.458} & \textbf{0.178} & \textbf{0.502} & \textbf{0.155} \\ \hline
MLLF\_B                 & \textbf{0.156} & \textbf{0.053} & \textbf{0.184} & \textbf{0.032} & \textbf{0.459} & \textbf{0.179} & \textbf{0.524} & \textbf{0.159} \\ \hline
MFP\_B                  & \textbf{0.154} & \textbf{0.055} & \textbf{0.191} & \textbf{0.030} & \textbf{0.455} & \textbf{0.179} & \textbf{0.502} & \textbf{0.156} \\ \hline
MLLFP\_B                & \textbf{0.155} & \textbf{0.055} & \textbf{0.183} & \textbf{0.031} & \textbf{0.468} & \textbf{0.206} & 0.550          & 0.207          \\ \hline
\end{tabular}

\caption{\label{tab:a5} 
RMSE of Prediction and Imputation of the curves for different methods under Non-Linear case $(f(X_i(t),t)=cos(X(t)^3*t)+5*t)$ when $n=500$ under different time points and sparsity settings.}
\end{table}

\end{document}